\documentclass{appolb}
\usepackage{epsfig}

\begin{document}
\title{Physics at LHC
}
\author{John~Ellis
\address{Theory Division, Physics Department, CERN, CH-1211 Geneva 23, Switzerland\\
~~\\
CERN-PH-TH/2006-239}
}
\maketitle
\begin{abstract}
The prospects for physics at the LHC are discussed, starting with the foretaste,
preparation (and perhaps scoop) provided by the Tevatron, in particular,
and then continuing through the successive phases of LHC operation. These include the start-up
phase, the early physics runs, the possible search for new physics in double diffraction, the
continuation to nominal LHC running, and the possible upgrade of the LHC luminosity.
Emphasis is placed on the prospects for Higgs physics and the search for supersymmetry.
The progress and discoveries of the LHC will set the time-scale and agenda for the
major future accelerator projects that will follow it.
\end{abstract}
  


\section{The Big Particle Physics Match} 

The principal goal of the LHC experimental programme is to explore directly for the first
a completely new scale of energies and distances, to the TeV scale and beyond. 
Among the specific objectives in this
exploration are the search for the Higgs boson(s), to look for whatever new physics may
accompany it, such as supersymmetry or extra dimensions, and (above all) find
something that the theorists did not predict. Many of these topics were discussed
extensively during this workshop, but this is just a concluding talk, not a summary of the
conference. Although I discuss the conference topics and refer to many of the talks, I confess that
most of the illustrations are taken from my own papers.

Since the World Football Cup has been very much in our minds during the workshop, I
use this as a metaphor for the big LHC particle physics match. We start in the training
camp provided by the Fermilab Tevatron collider, B factories, new theoretical ideas and
calculations. Then we warm up with a brief discussion of LHC installation and commissioning,
and the prospects for the pilot run planned for late 2007. Moving into the first half of the
match, we discuss the prospects for the first 1 to 30~fb$^{-1}$. Then, in injury time we
discuss the prospects for diffractive Higgs production at the LHC. In the second half, we
discuss the prospects for further LHC running up to the 300~fb$^{-1}$ foreseen for the
full LHC experimental programme at the design luminosity. Inevitably, there are plans
for extra time with the possibility that the LHC luminosity could be upgraded to
$10^{35}$~cm$^{-2}$s$^{-1}$ (the SLHC project). Finally, what are the prospects for the 
penalty shoot-out to decide what happens in the next round of particle physics
experiments: ILC vs CLIC vs DLHC vs TLHC (projects for doubling or tripling the LHC
energy)?

\section{Training Camp}

The Tevatron provides access to many topics in Standard Model physics that will also
play a key role at the LHC, such as QCD, B physics, electroweak physics, top physics
and the search for the Higgs boson. Indeed, the first evidence for the Higgs boson may
well be provided by the Tevatron. The current upper limits on Higgs production at the
Tevatron are already within about an order of magnitude of the Standard Model explanation,
depending on its mass~\cite{Gonzalez}. The sensitivity achieved so far corresponds approximately to
the expectations for the amount of luminosity already accumulated, and it is hoped to
accumulate up to an order of magnitude more data than have been analyzed so far.

In the mean time, the Tevatron is making progress with several fundamental electroweak
measurements. The large samples of $W$ bosons provide an opportunity to measure
more accurately the $W$ mass, whose dominant  errors are due to uncertainties in the
parton distribution functions and the lepton energy scale. Beyond single $W$ and $Z$
production, $WW$ pair production has been observed at the $5 - \sigma$ level, and
associated $WZ$ production at the $3.3 - \sigma$ level~\cite{Hays}. Detailed understanding of $W$
production will be needed for both the Tevatron and the LHC: events with the $W$ 
accompanied by jets will
be very important backgrounds for many new physics channels, for example to the search
for Higgs production in association with the $W$.

The Tevatron experiments have made good progress in understanding QCD effects in events 
with photons and jets~\cite{Zielinski}. QCD not only provides the most 
significant backgrounds for many new processes,
but is also important in its own right, and as a tool for discovering new physics. It will
be necessary to study not only the jet total cross sections but also their azimuthal angular
distributions and correlations: one should be able to describe all the corners of the multijet 
phase space, where new physics may lurk.

The Tevatron experiments are making rapid progress in top physics~\cite{Zielinski}. 
They have established that
it has charge $+ 2/3$ and spin $1/2$. The mass has been measured with impressive
accuracy:
\begin{equation}
m_t \; = \; 171.4 \pm 2.1~{\rm GeV}.
\label{topmass}
\end{equation}
Taken together with the measurement of $m_W$, this measurement favours a relatively
light Higgs mass. An interesting development is the observation of top and $W$ peaks
in all-jet event samples. One may hope that, in the future at the LHC, it will be possible 
at the LHC to reconstruct new particles via their multijet decays.

The B factories~\cite{Bartoldus} and the Tevatron~\cite{Qian} are also making important 
strides in the study of B physics.
It has been possible to predict the amount of CP violation in B decays on the basis of
CP-conserving observables. These predictions have been confirmed, indicating that the
Cabibbo-Kobayashi-Maskawa (CKM) mechanism is the dominant source of CP violation
in B physics. However, one may still wonder whether CKM is the whole story. The
measurement of $\sin 2 \beta$ are perhaps not in perfect agreement with predictions based on
other measurements, and there are still puzzles in the comparison between the values of
$\sin 2 \beta$ found in $B \to J/\psi K$ and other decay modes~\cite{Bigi}. It will be necessary to
improve the present measurements of the other angles $\alpha, \gamma$ of the unitarity
triangle. Meanwhile, the rare decays $b \to s \gamma, s \ell^+ \ell^-$ are providing
interesting windows on new physics. They, together with the exciting recent measurement
of $B_s$ mixing and the detection of $B \to \tau \nu$ decay, start to exert significant
pressure on supersymmetry, for example.

What is happening meanwhile in the theoretical training camp? Ways to simplify dramatically
certain QCD calculations have recently emerged from string theory and the use of
twistor techniques~\cite{Lykken}. However, these have not yet led to large increases in the 
range of QCD
and electroweak processes calculated to NLO or beyond. A large fraction of the processes
that are potentially important backgrounds to new physics searches at the LHC remain
uncalculated at the NLO level, including the following final states: $VV +$ one jet, $H + 2$
jets, ${\bar t} t + {\bar b}b$, ${\bar t}t + 2$ jets, $VV + {\bar b}b$, $VV + 2$ jets, $V + 3$ jets,
$VVV$. 

The Higgs search is particularly stimulating for theoretical developments~\cite{Harlander} in such calculations,
such as phase space and higher-order Monte Carlos. Higher orders are essential: for
example, one knows that
\begin{equation}
\sigma(gg \to H) \; \sim \; \sigma_{LO}(1 + 0.7 + 0.3 + ...) \; \sim \; 2 \sigma_{LO}.
\label{Higgs}
\end{equation}
The clear understanding of theory will open new experimental windows, for example in the
measurements of Higgs couplings and probes of CP properties. These calculations also need to new conceptual understanding, for example of the bottom-quark density in the proton and higher
orders in supersymmetry. Theorists foresee exciting times ahead: the (N)NLO era at hadron
colliders has begun, and they are looking forward to Higgs physics with data!

In parallel with this preparatory work within the Standard Model, theorists are naturally thinking
busily about open questions beyond the Standard Model~\cite{Pokorski}. Is the origin of particle masses
really due to a Higgs boson, and is it accompanied by some other new physics? Why are
there so many types of matter particles, and is this linked with the observed matter-antimatter
difference as in the CKM mechanism? Are the different fundamental forces unified and, if so, does
unification really occur at very high energy $\sim 10^{16}$~GeV, as in popular models? How
are quantum theory and general relativity reconciled, perhaps via string theory, perhaps with
large extra space dimensions? The LHC may be able to illuminate all these issues, either
directly or indirectly. 

In my opinion, the most plausible extension of the Standard Model is
supersymmetry~\cite{Kalinowski}, for the following reasons.

$\bullet$ Supersymmetry may accompany the Higgs boson and play a key role in stabilizing the
magnitude of its mass, and hence the electroweak symmetry breaking scale. As is well known,
the leading loop corrections to the squared Higgs mass diverge quadratically. Since boson
and fermion loops have opposite signs, these quadratic divergences may be cancelled if a
specific relation between the fermion and boson couplings holds, namely that embedded in
supersymmetric models~\cite{hierarchy}.

$\bullet$ Supersymmetry facilitates the unification of the fundamental forces: extrapolating
the strengths of the strong, weak and electromagnetic interactions measured at low energies
does not give a common value at any energy in the absence of supersymmetry, but there is a
common value at an energy $\sim 10^{16}$~GeV in the presence of supersymmetry~\cite{unification}.

\begin{figure}[htb]
\includegraphics[width=.45\textwidth,height=5.4cm]{mhplot_2006a.eps}
\includegraphics[width=.45\textwidth,height=5.4cm]{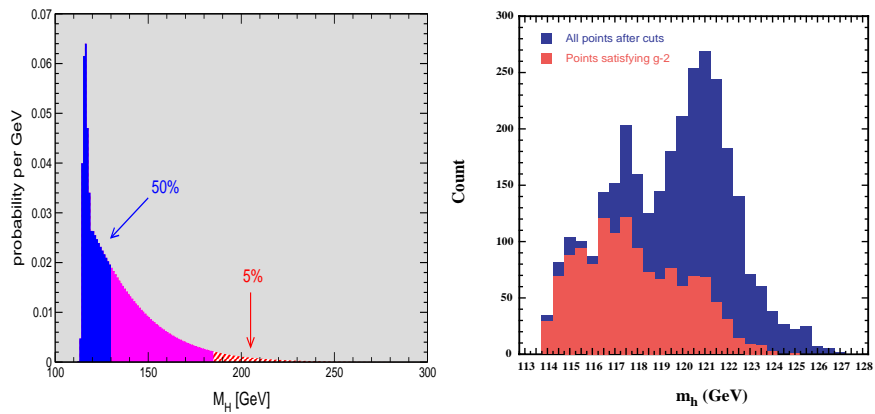}\\[1em]
\caption{\it (a) The probability distribution for the mass of the Standard model Higgs boson
found~\protect\cite{Erler} by combining the direct LEP search information with an analysis of
the precision electroweak data, and (b) a histogram of the mass of the lightest Higgs boson $h$
found in a sampling of CMSSM parameters, both without (blue) and with (red) the inclusion of
$g_\mu - 2$.}
\label{fig:lightH}
\end{figure}

$\bullet$ Supersymmetry predicts a low mass for the Higgs boson, probably below 
130~GeV~\cite{higgs},
as suggested by a global fit to precision electroweak data~\cite{Erler}, as seen in
Fig.~\ref{fig:lightH}.

$\bullet$ Supersymmetry provides a natural candidate for the cold dark matter advocated by
astrophysicists~\cite{EHNOS}.

In the latter case, the lightest supersymmetric particle (LSP) should have neither strong nor
electromagnetic interactions, since otherwise it would bind to conventional matter and be
detectable as an apparent anomalous heavy nucleus. {\it A priori}, possible weakly-interacting
scandidates included sneutrinos, but these have now been excluded by LEP and direct
searches. Nowadays, the scandidates most considered are the lightest neutralino $\chi$
and (to a lesser extent) the gravitino (which would be a nightmare for astrophysical detection, 
but not necessarily
for collider experiments, as discussed below).

\begin{figure}[htb]
\centerline{\includegraphics[height=2.5in]{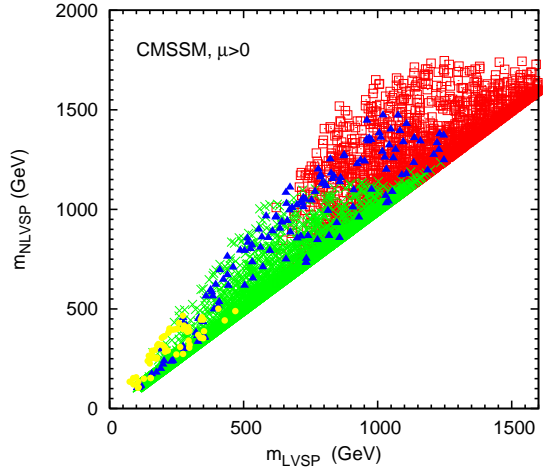}}
\caption{\it A scatter plot of the masses of the lightest visible supersymmetric particle (LVSP)
and the next-to-lightest visible supersymmetric particle (NLVSP) found in a sampling of
CMSSM parameters (red), including those that produce a suitable amount of
dark matter (blue), most of which are detectable at the LHC (green), but perhaps not directly as
astrophysical dark matter (yellow)~\protect\cite{EOSS}.}
\label{fig:heavy}
\end{figure}

Assuming that the LSP is the lightest neutralino, the parameter space of the constrained 
minimal supersymmetric extension of the Standard Model (CMSSM) is restricted by the need 
to avoid a stau LSP, by the measurements of $b \to s \gamma$ that agree with the Standard Model,
by the range of cold dark matter density allowed by WMAP and other observations, and by the
measurement of $g_\mu - 2$. These requirements are consistent with relatively large masses
for the lightest and next-to-lightest visible supersymmetric particles, as seen in 
Fig.~\ref{fig:heavy}~\cite{EOSS}.
We see there that most of the models that provide cosmological dark matter are detectable at the 
LHC, though this is not guaranteed, whereas the dark matter is directly detectable in only
a rather smaller fraction of models.

\begin{figure}[htb]
\includegraphics[width=.45\textwidth,height=5.4cm]{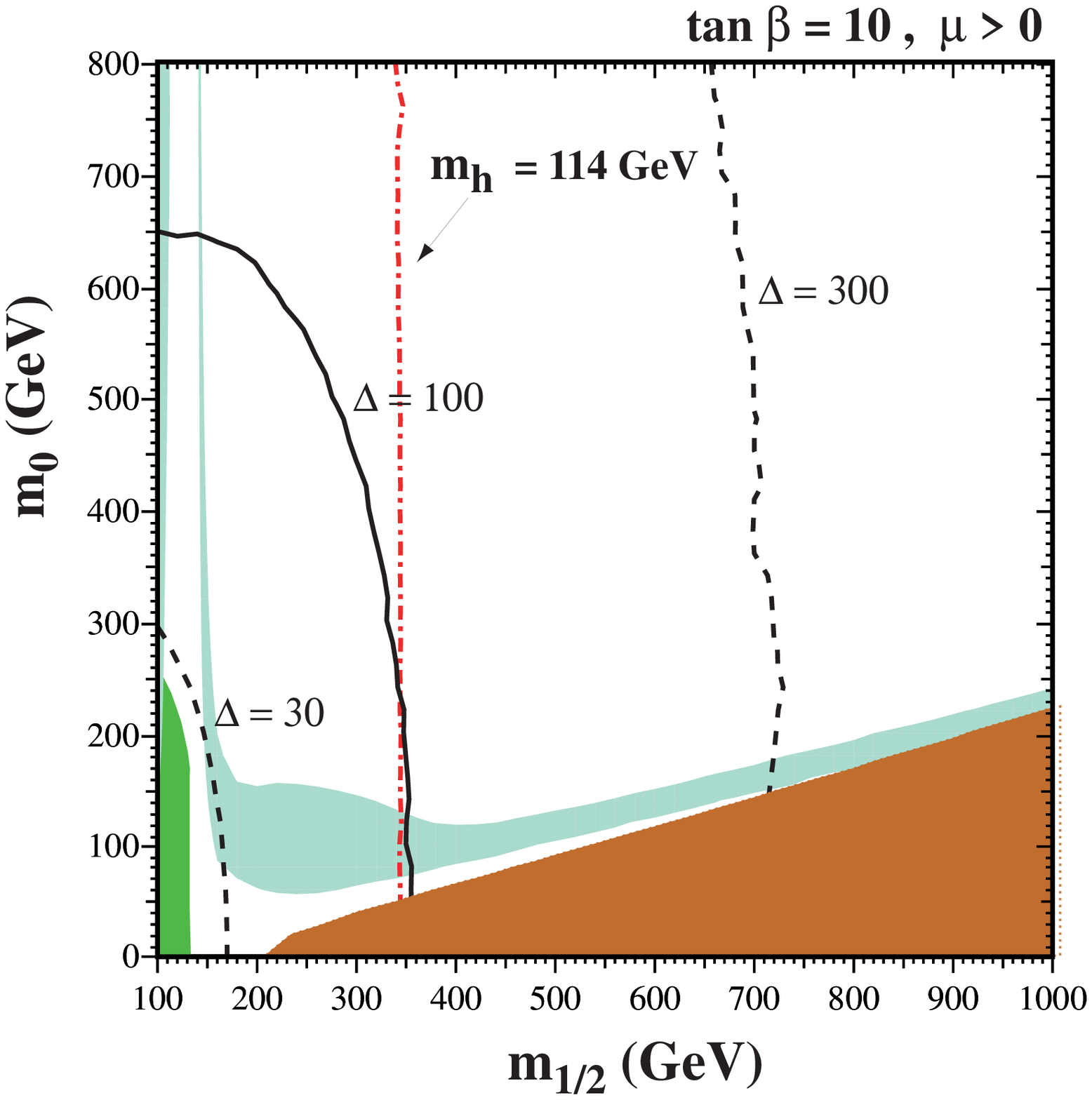}
\includegraphics[width=.45\textwidth,height=5.4cm]{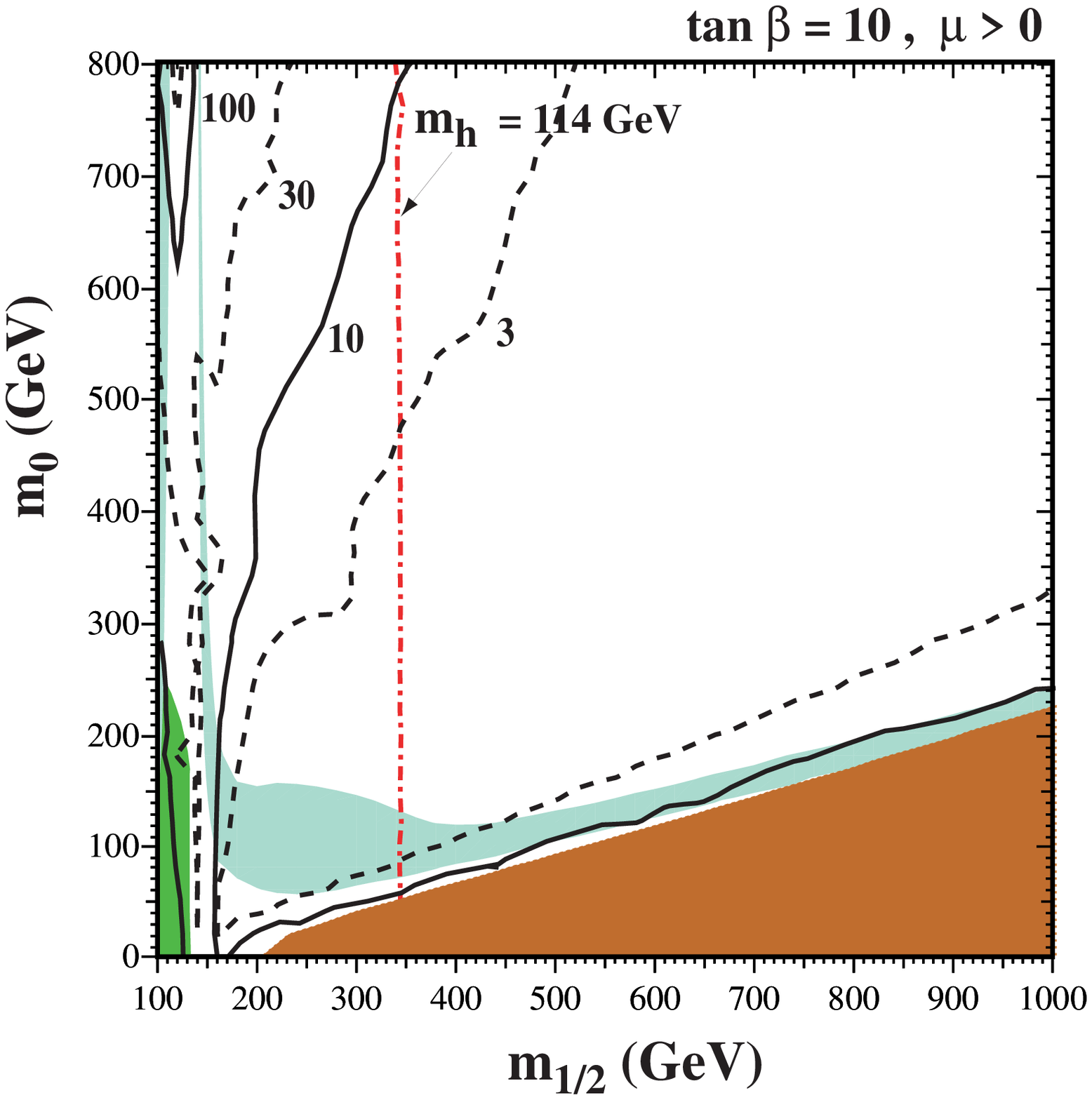}\\[1em]
\caption{\it The amounts of fine-tuning required for the electroweak mass scale (left) and in
order to obtain the appropriate cold dark matter density (right)~\protect\cite{EOS}.}
\label{fig:finetune}
\end{figure}

As discussed at this meeting,
larger sparticle masses generally require more fine-tuning in order to obtain the appropriate
electroweak symmetry-breaking scale, as seen in Fig.~\ref{fig:finetune}(a)~\cite{EOS}. Larger sparticle
masses also imply that the underlying supersymmetric mass parameters must be adjusted
more accurately in order to obtain the appropriate cold dark matter density, but this effect is not
very strong, as seen in Fig.~\ref{fig:finetune}(b)~\cite{EOS}. It is difficult to know how much significance to attach to the absolute amount
of fine-tuning, which depends on the definition of the measure of fine-tuning and on one's pain
threshold. Personally, I do not find the level of fine-tuning imposed by the present limits particularly
painful.

\begin{figure}[htb]
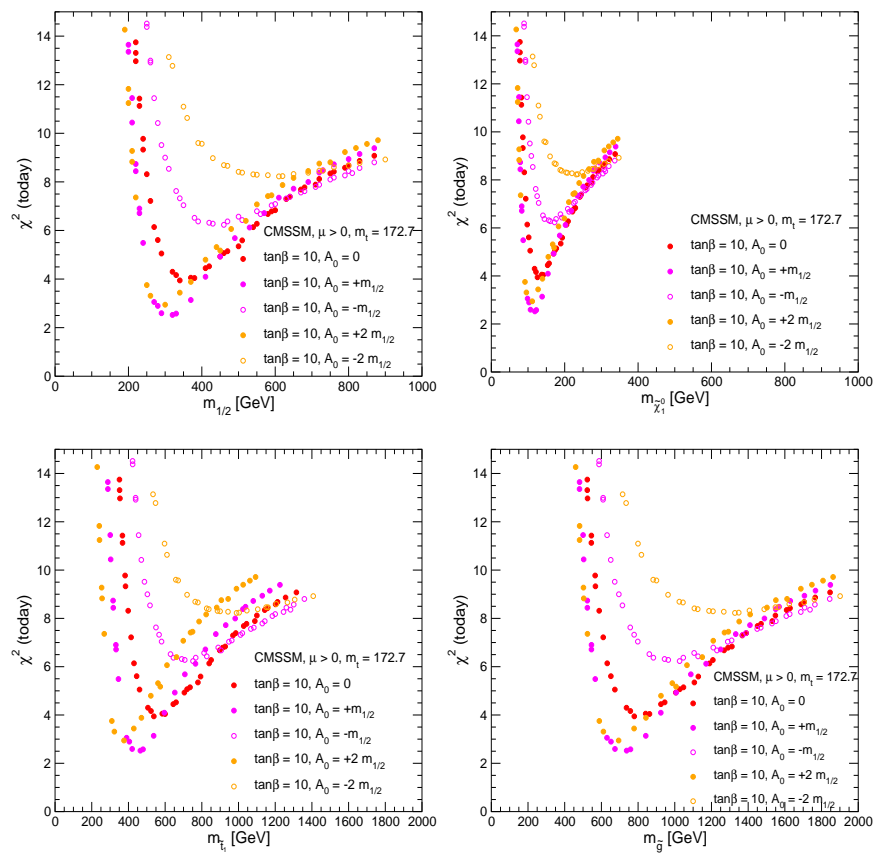

\includegraphics[width=.45\textwidth,height=5.4cm]{ehow.CHI11a.1727.cl.eps}
\includegraphics[width=.45\textwidth,height=5.4cm]{ehow.mass11a.1727.cl.eps}\\[1em]
\includegraphics[width=.45\textwidth,height=5.4cm]{ehow.mass19a.1727.cl.eps}
\includegraphics[width=.45\textwidth,height=5.4cm]{ehow.mass23a.1727.cl.eps}
\caption{\it The $\chi^2$ distributions found for (a) $m_{1/2}$, (b) $m_\chi$, 
(c) $m_{\tilde t_1}$ and (d) $m_{\tilde g}$ in an analysis of the CMSSM
parameter space including precision electroweak data and $g_\mu - 2$~\protect\cite{EHOW}.}
\label{fig:EHOW}
\end{figure}

Within the overall range allowed by the experimental constraints, are there any hints what the
supersymmetric mass scale might be? The high-precision measurements of $m_W$ and the
weaK mixing angle $\sin^2 \theta_W$ each favour a relatively small sparticle mass scale. On the
other hand, the rate for $b \to s \gamma$ shows no evidence for light sparticles, and the experimental
upper limit on $B_s \to \mu^+ \mu^-$ begins to exclude very small masses. The strongest indication
for new low-energy physics, for which supersymmetry is just one possibility, is offered by $g_\mu - 2$,
if one uses $e^+ e^-$ data to calculate the Standard Model contribution. Putting this together with
the other precision observables, one finds a preference for light sparticles, which is more
pronounced for $\tan \beta = 10$ than for $\tan \beta = 50$~\cite{EHOW}, as seen in 
Fig.~\ref{fig:EHOW}.

There are alternative scenarios for new low-energy physics beyond the Standard 
Model~\cite{Pokorski}, such as
one or more large extra dimensions. However, such models must steer a narrow course between
the Scylla of excessive effects in electroweak data and the Charybdis of inaccessibility at the LHC.
That would be a shame, as extra dimensions offer a rich alternative to the phenomenology of
supersymmetry. Another rich alternative is offered by little Higgs models, which feature new fermions,
gauge bosons and Higgs bosons that might be accessible to the LHC. Using modern duality
ideas, these can actually be regarded as deconstructed versions of higher-dimensional models.

Only the LHC will be able to tell us which, if any, of these theoretical speculations has any connection
to reality.

\section{Warming up}

Perhaps this section should better be called `cooling down'. The troubles of the LHC cryogenic system are now behind us, most of the magnets have already been installed in the LHC tunnel, and they are
now been connected together. When the machine is closed, somewhat after the middle of 2007,
commissioning of the machine will start in earnest, and a pilot run at the LHC injection energy of
450~GeV per beam is planned for November 2007~\cite{Engelen}. This will permit the debugging of the
accelerator and its detectors~\cite{Jenni,Dobrzynski}. They will get a look at some
minimum-bias events and a few moderately low-energy jets. The commissioning of the machine 
will then be
completed in the first part of 2008, and there will then be running at the design energy of 7~TeV
per beam.

\section{First Half}

The first step in LHC physics will be to measure and understand minimum-bias events, which
will proceed in parallel with understanding the detectors. The next physics to be measured will
be jets, which will be compared with QCD predictions. During this phase, a key task will be the
energy calibrations of the detectors. The next Standard Model processes to be measured and
understood will be the $W$ and $Z$, which will permit the lepton energies to be calibrated.
The following Standard Model milestone will be top physics, which will provide the
opportunity to calibrate better the jet energies and the resolution in missing $E_T$. 
Only after these steps will the search for the Higgs boson be able to start in earnest. This will
require combining many signatures in different channels, and will entail excellent
understanding of the detectors. The Higgs will not jump out in the same way as did the $W$ and 
$Z$, or even the top quark. Around the time that Higgs searches get underway, the first
searches for supersymmetry or other new physics beyond the Standard Model will also start.

What are the prospects for Standard model studies at the LHC~\cite{Boonekamp}? 
The first necessity will be
to obtain the jet multiplicity distributions from data: at present there is disagreement among the
available Monte Carlo programmes. Also important for the $W$ and $Z$ measurements themselves, as well as for subsequent searches for new physics, will be the quantitative understanding of jet activity
in $W$ and $Z$ production. Theoretical uncertainties are likely to dominate the
statistical experimental errors expected for $W$ and $Z$ physics. The conventional view has
been that the calibration of the lepton energy scale, the understanding of accompanying hadronic
activity in $Z$ production and measurements of the parton distribution functions will enable the
error in $m_W$ to be reduced below 10~MeV. It was suggested here that the use of off-shell
$Z$ production to reduce extrapolation errors might help reduce the error in $m_W$ to be
reduced below 5~MeV. Sceptics should remember that many of the final experimental arrors
achieved at LEP were far smaller than had been expected before accelerator operations
started!

Likewise, it may be possible to reduce the experimental error in $m_t$ below 1~GeV. This
would require detailed understanding of background event pile-up, the jet energy scale and the underlying events, but the error may eventually be reduced below $\sim 0.5$~GeV. The combination
of precise measurements of $m_W$ and $m_t$ will enable the Higgs mass to be estimated with
high precision within the Standard Model. The confrontation with the direct measurement will
provide an important check on radiative corrections, and perhaps to
look for beyond the Standard Model such as superysmmetry.

As already mentioned, in the search for new physics at the LHC, the primary task will be to
understand the Standard Model and the detectors, which will require
detailed studies of calibrations, alignment and systematics. Only when the 
most important Standard Model cross sections are understood will it be possible to
look for signatures of new physics beyond the Standard Model. One sometimes hears talk of 
looking for the
signatures of specific scenarios such as supersymmetry, at other times of
signature-based searches, e.g., for monojets. In practice, this will be a false
dichotomy: one will look for generic signatures of new physics that could be due to
several different new-physics scenarios. For example, missing-energy events could be due
to supersymmetry, extra dimensions, black holes or the radiation of gravitons into
extra dimensions. If one finds such events, the challenge will be to distinguish between the
different scenarios. In the specific case of distinguishing between suspersymmetry and universal extra
dimensions, various tools will be available. The spectra of higher excitations would be
different in the two scenarios, the different spins of particles in cascade decays would
yield distinctive spin correlations, and the spectra and asymmetries of, e.g., dileptons,
would be distinguishable.

\begin{figure}[htb]
\centerline{\includegraphics[height=2.5in]{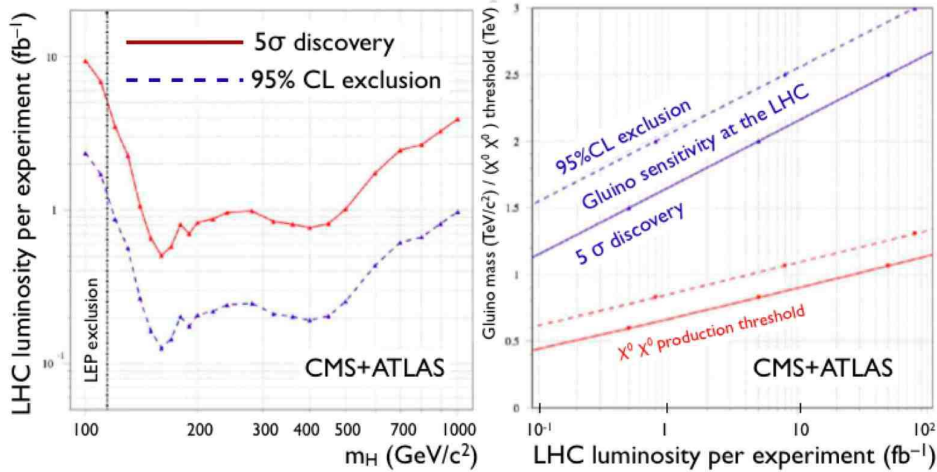}}
\caption{\it The combined ATLAS and CMS sensitivity as a function of the LHC luminosity to (a) 
a Standard Model Higgs boson and (b) the gluino~\protect\cite{POFPA}. 
In the latter case, we also show the corresponding
threshold for $e^+ e^- \to \chi \chi$. }
\label{fig:gluino}
\end{figure}

What is the discovery potential of this initial LHC running? As seen in Fig.~\ref{fig:gluino},
a Standard Model Higgs boson could be discovered with 5-$\sigma$ significance with
5~fb$^{-1}$ of integrated and well-understood luminosity~\cite{Unal}, whereas 1~fb$^{-1}$ would
already suffice to exclude a Standard Model Higgs boson at the 95~\% confidence
level over a large range of possible masses~\cite{POFPA}. However, this Higgs signal would receive
contributions from many different decay signatures, such as $\tau \tau$, $\gamma \gamma$,
${\bar b}b$, $WW$ and $ZZ$. Finding the Higgs boson will require understanding the
detectors very well so as to find each of these signatures with good efficiency and low background.
Therefore, the Higgs discovery announcement may not come very soon after the accelerator
produces the required integrated luminosity!

Paradoxically, some new physics scenarios may be easier to spot, if their mass scale is not
too high, such as supersymmetry~\cite{Spiropulu}. For example, as seen in Fig.~\ref{fig:gluino}, 
0.1~fb$^{-1}$ of luminosity should be 
enough to detect the gluino
at the 5-$\sigma$ level if $m_{\tilde g} < 1.2$~TeV, and to exclude its existence below 1.5~TeV
at the 95~\% confidence confidence level~\cite{POFPA}. This amount of 
integrated luminosity could be gathered with an
ideal month's running at 1~\% of the design instantaneous luminosity. An integrated
luminosity of 10~fb$^{-1}$, such as could be obtained by 2009, would suffice to
discover the gluino if it weighs less than 2.2~TeV, and to exclude it below 2.6~TeV, as seen
in Fig.~\ref{fig:gluino}.

If the neutralino is the LSP, all other sparticles must be heavier, and hence the threshold for 
producing sparticle pairs in $e^+ e^-$ collisions must exceed $2 m_\chi$. In supersymmetric 
models with universal gaugino masses that unify at the GUT scale,
the mass of the lightest neutralino is proportional to the gluino mass. Hence the gluino mass
determines the $e^+ e^-$ threshold~\cite{POFPA}, as also seen
in Fig.~\ref{fig:gluino}.

As can be seen from Fig.~\ref{fig:gluino}, almost immediately after being turned on, the
LHC will be able to tell us whether sparticles can be pair-produced at the initial centre-of-mass
energy of the ILC, namely 0.5~TeV. With 10~fb$^{-1}$, such as might be analyzed by 2010,
one would very likely know whether sparticles could be pair-produced at the ultimate ILC
centre-of-mass energy of 1~TeV. If not, a higher-energy $e^+ e^-$ collider such as CLIC would
be desirable. Even if some sparticles are light enough to be produced at the ILC,
CLIC would certainly be able to study more sparticle species.
 
In parallel, heavy-ion collisions at the LHC may resolve some of the puzzles~\cite{Salgado} 
posed by the RHIC data~\cite{Roland}. 
These indicate that the initial-state parton distributions may be saturated, and that
there is very rapid thermalization. After this, a fluid with very low viscosity and large transport
coefficients seems to be produced. One of the surprises was that the quark-gluon (?) medium
produced at RHIC seems to be strongly-interacting. The final state exhibits jet quenching and the
semblance of cones of energy deposition akin to Machian shock waves or {\v C}erenkov
radiation patterns, indicative of very fast particles moving through a medium with a lower
speed of sound or light.

A new idea that may explain parton saturation is the colour-glass condensate (CGC)~\cite{CGC},
which gives qualitative understanding of hadron multiplicities in some kinematic regions,
and may also lead to rapid thermalization. The AdS/CFT correspondence provides a hint how
the low viscosity may arise, and also provides a scheme for calculating jet quenching and the
propagation of heavy quarks through the strongly-interacting `plasma'~\cite{AdSCFT}.

Experiments at the LHC will enter a new range of temperatures and pressures~\cite{Nappi}, 
thought to be far
into the quark-gluon plasma r{\' e}gime. Parton saturation effects are expected to be more
significant, and it may be possible to subject the CGC to incisive experimental tests. A real phase transition between the hadronic and
quark-gluon descriptions is not expected, it is more likely to be a cross-over that may not have a
distinctive experimental signature at high energies. However, it may well be possible at the LHC
to see quark-gluon matter in its weakly-interacting high-temperature phase. The larger
kinematic range should also enable ideas about jet quenching and radiation cones to be
tested.

The fourth major LHC experiment, LHCb~\cite{Nakada}, will be able to compare flavour physics and CP
violation at the tree level, where the CKM model seems to work very well, and at loop level,
where new physics may be relatively more important. There are still some puzzles: for instance,
the value of $\beta$ found in $B \to J/\psi K$ does not agree perfectly with predictions
based on previous measurements, and the values found in other $B$ decays may be slightly
different. LHCb will be able to measure subtle CP-violating effects in $B_s$ decays, and will also
improve measurements of all the angles $\alpha, \beta$ and $\gamma$ of the unitarity triangle.
The LHC will also provide high sensitivity to rare $B$ decays, which may open another
window on CP violation beyond CKM.

Since the $B$ factory experiments have established that the CKM mechanism is dominant, 
the question is no longer whether the CKM model is `right'. The task is rather to look for additional sources of CP
violation, which must surely exist, e.g, in order to create the cosmological matter-antimatter
asymmetry via baryogenesis. It is an open question whether these may provide new physics at the
TeV scale accessible to the LHC. On the other hand, if the LHC does observe any new physics,
such as the Higgs boson and/or supersymmetry, it will become urgent to understand its flavour
and CP properties.

\section{Injury Time}

An interesting option for extending the LHC experimental programme already in this
initial phase is to look for new physics such as the Higgs boson in diffractive scattering~\cite{Bartels}.
The cross section for producing the Standard Model Higgs boson exclusively via double 
diffraction is large enough for its observation to appear possible with a luminosity of
$10^{33}$~cm$^{-2}$s$^{-1}$~\cite{deRoeck}, at least if its mass is in the lower part of the range 
suggested by the
precision electroweak data. The cross section may even be enhanced in some
extensions of the Standard Model such as supersymmetry, and the production
mechanism offers novel opportunities to look for CP-violating Higgs couplings and mixing
effects.

The existing ATLAS and CMS detectors are not sensitive to the double-diffractive
production of low-mass particles, and nor is TOTEM~\cite{Avati}. Greater sensitivity to low-mass
Higgs production, in particular, could be obtained by adding detectors to either ATLAS
and/or CMS in far-forward positions 420~m from the interaction points, able to detect
leading protons that have lost small fractions of the beam momenta. These would need
to rely on triggers on activity in the central detectors, as information arriving from 
$\pm 420$~m would be too late to be included in the trigger. An FP420 R\&D project is
underway with support from both ATLAS and CMS~\cite{deRoeck}.

\section{Second Half}

The nominal LHC luminosity is $10^{34}$~cm$^{-2}$s$^{-1}$, and one may hope to
accumulate 300~fb$^{-1}$ after several years of running at this design luminosity. What
can one hope to learn with such an integrated luminosity?

\begin{figure}[htb]
\centerline{\includegraphics[height=2.5in]{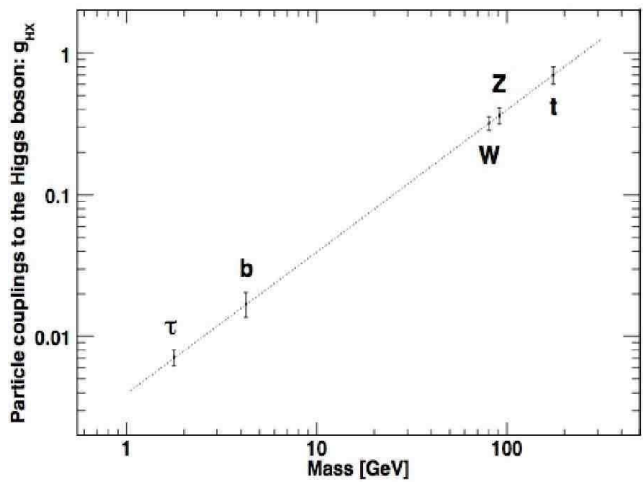}}
\caption{\it The potential of the LHC for determining the Higgs couplings to different Standard
Model particles, as a function of their masses~\protect\cite{POFPA}.}
\label{fig:Higgscouplings}
\end{figure}

Much may be learned about the properties of the Higgs boson, depending on its mass.
If the Higgs is observed to decay into either $\gamma \gamma$ or $ZZ$, one will know that
it cannot have spin 1, and observations of angular distributions and correlations in $Z Z^{(*)}$
decays will enable the spin and CP properties of the Higgs to be 
determined~\cite{Higgsspin,HiggsCP}. 
It will also be
possible to measure even invisible Higgs decays at the 15 to 30~\% level. Overall, it will be
possible to measure many Higgs-particle couplings (to $\tau, b, W, Z$ and $t$) at the 10 to 
20~\% level, enabling one
to check whether they are proportional to the particle masses and hence whether the
Higgs boson really accomplishes its assigned task of providing their masses~\cite{POFPA},
as seen in Fig.~\ref{fig:Higgscouplings}.

\begin{figure}[htb]
\centerline{\includegraphics[height=5.5in]{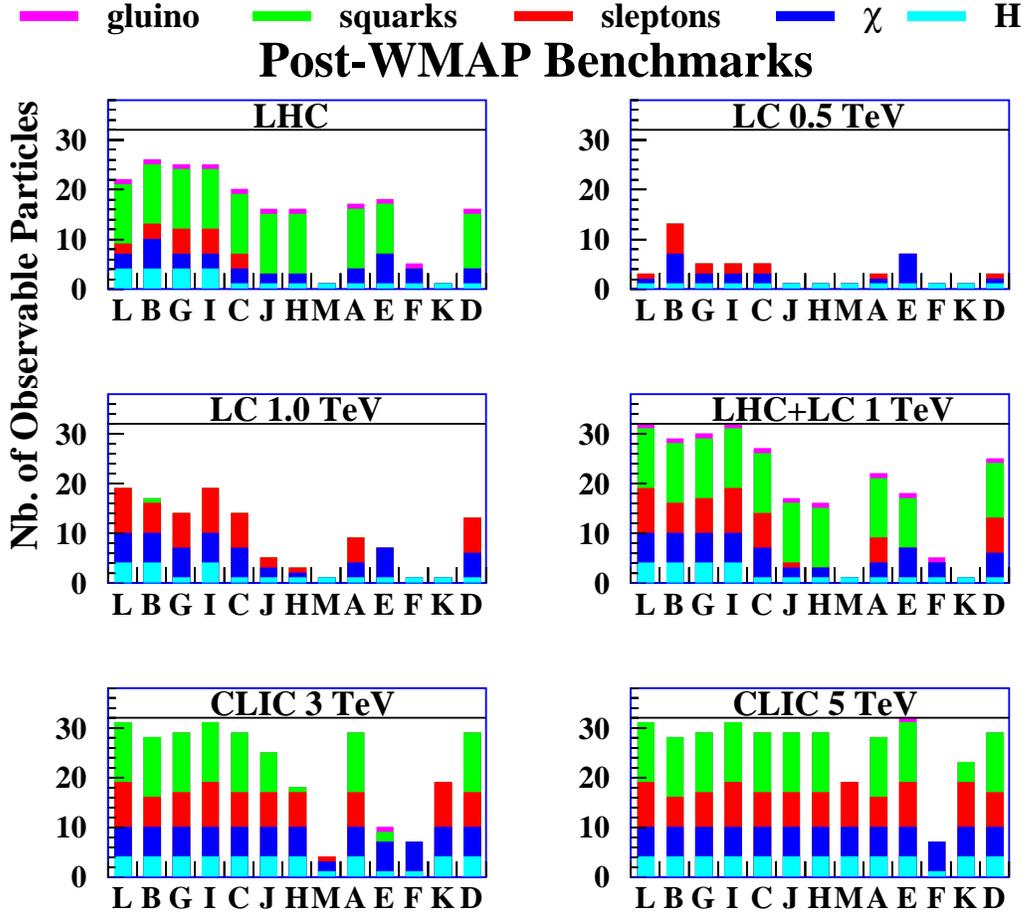}}
\caption{\it The numbers of different sparticle species that could be detected at the LHC and
$e^+ e^-$ colliders with different centre-of-mass energies, in various benchmark supersymmetric
scenarios with universal soft supersymmetry-breaking masses~\protect\cite{Bench2}.}
\label{fig:Bench2}
\end{figure}

The full LHC reach for supersymmetry will extend to $m_{\tilde g} \sim 3$~TeV, enabling the
LHC to discover supersymmetry over most of the range where it can provide dark matter,
and, if it is not found, forcing most theorists to lose faith in its relevance to the mass problem.
Since the LHC generally produces heavier sparticles that decay through cascades into the
LSP, in many supersymmetric scenarios it will provide opportunities to measure the masses
and other properties of a number of different sparticle species, as seen in
Fig.~\ref{fig:Bench2}~\cite{Bench1,Bench2}.

As already mentioned, it has normally been assumed that the LSP is the lightest neutralino $\chi$, 
but the gravitino ${\tilde G}$ has recently been attracting increased attention.
The ${\tilde G}$ would be a nightmare for direct searches for astrophysical dark matter, 
because of its extremely weak, gravitational-strength
couplings to ordinary matter, but this
feature also means that it could be a bonanza for the LHC~\cite{Bench3,Are}, as we now discuss.

The gravitational-strength interactions of the ${\tilde G}$ imply that the next-to-lightest
supersymmetric particle (NLSP) would have a very long lifetime for gravitational decay~\cite{GDM}. 
For example,
if the NLSP is the lighter stau slepton ${\tilde \tau}$, its decay rate would be
\begin{equation}
\Gamma_{{\tilde \tau} \to \tau {\tilde G}} \; = \; \frac{1}{48 \pi M_P^2}
\frac{m_{\tilde \tau}^5}{m_{\tilde G}^2} \left( 1 - \frac{m_{\tilde G^2}}{m_{\tilde \tau}^2} \right)^4 .
\label{NLSPlife}
\end{equation}
This renders the ${\tilde \tau}$ metastable on the scale of LHC detectors,
and could be measurable in hours, days, weeks, months or conceivably years!

\begin{figure}[htb]
\centerline{\includegraphics[height=2.5in]{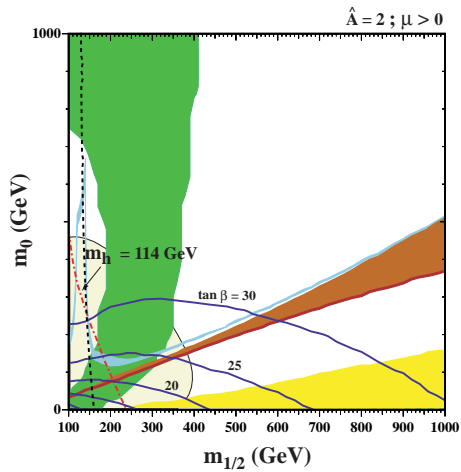}}
\caption{\it The $(m_{1/2}, m_0)$ plane in a mSUGRA scenario with the trilinear soft
supersymmetry-breaking parameter ${\hat A} = 2$. The green regions are excluded by 
$b \to s \gamma$, the brick-red regions because the LSP would be charged, and the yellow regions
because of the cosmological light-element abundances~\protect\cite{GDM}.}
\label{fig:mSUGRA}
\end{figure}

Generic possibilities for the NLSP include the $\chi$ and the ${\tilde \tau}$. Both of these
possibilities are generic, but they are strongly constrained by astrophysics and 
cosmology~\cite{EOV,CEFOS}.
In the CMSSM, the relation between $m_{\tilde G}$ and the other sparticle masses is left
unspecified, and in $\chi$ LSP scenarios it is implicitly assumed to be quite heavy. However,
in minimal supergravity (mSUGRA) scenarios $m_{\tilde G} = m_0$ at some input scale, and
there are additional constraints between the soft tri- and bilinear couplings: $B = A - 1$~\cite{GDM}. 
As a
result, one knows for each point in the $(m_{1/2}, m_0)$ plane which particles are the LSP
and NLSP, as well as the value of $\tan \beta$, as seen in Fig.~\ref{fig:mSUGRA}~\cite{GDM}.

\begin{figure}[htb]
\centerline{\includegraphics[height=2.5in]{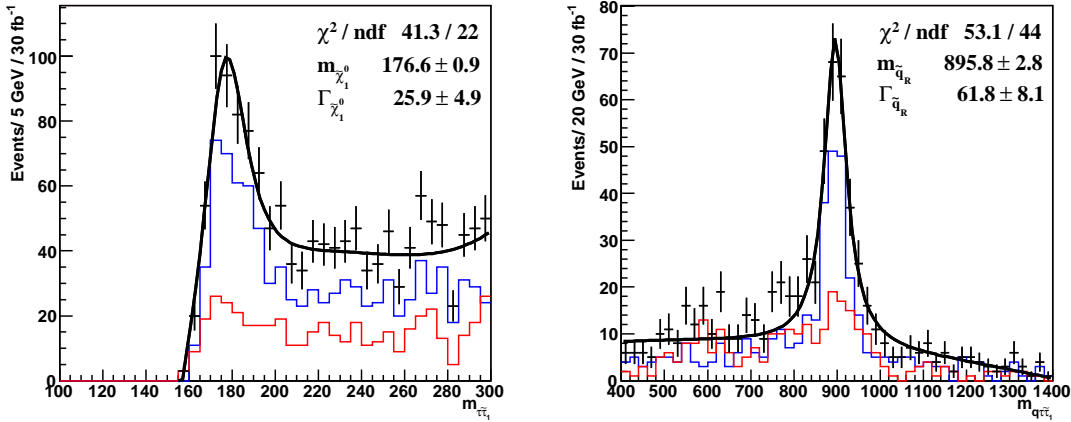}}
\caption{\it The reconstruction of heavier sparticles decaying into the ${\tilde \tau}$ in a scenario where it is the NLSP and the gravitino is the LSP~\protect\cite{Are}.}
\label{fig:Are}
\end{figure}

\begin{figure}[htb]
\centerline{\includegraphics[height=4.5in]{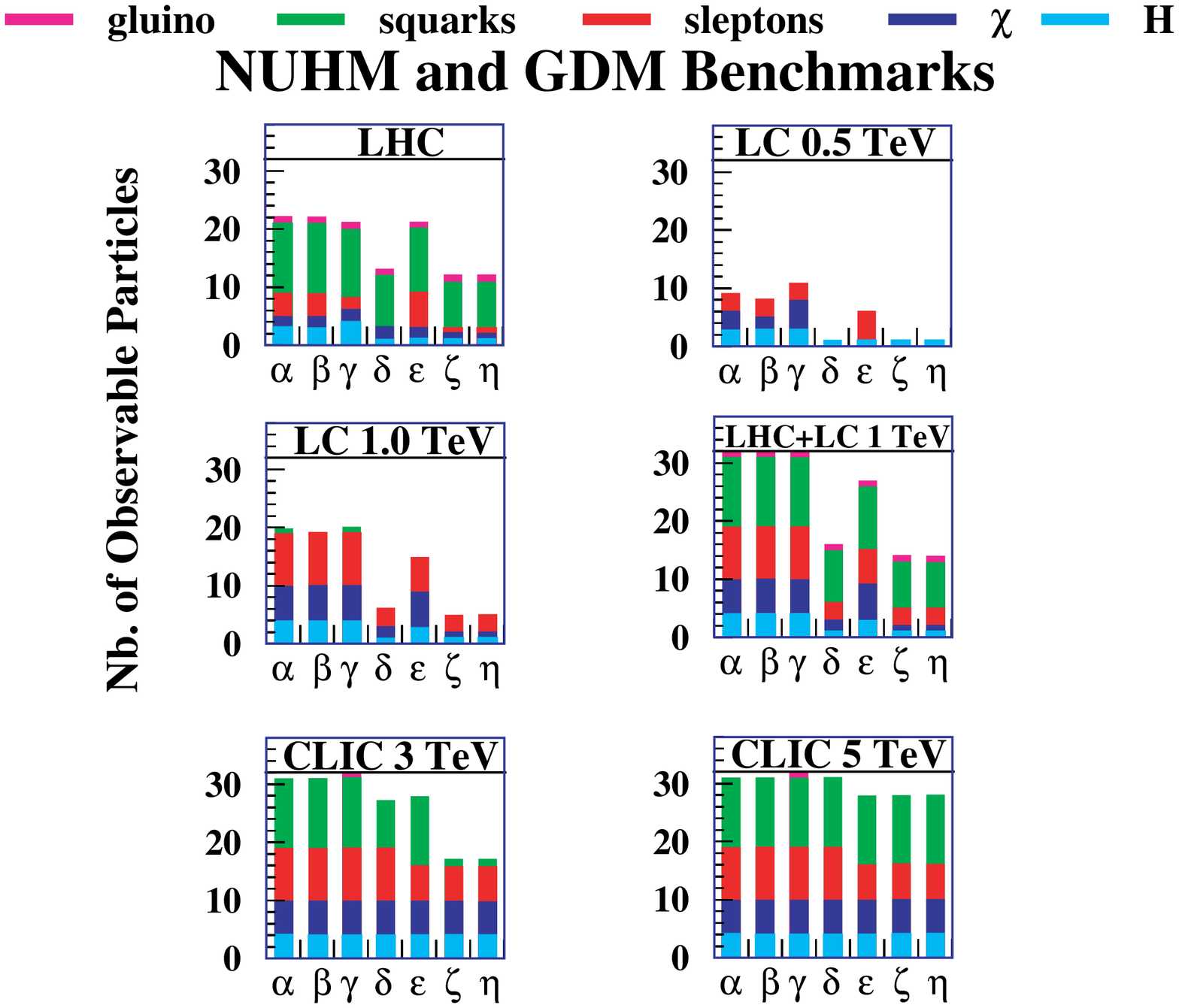}}
\caption{\it The numbers of different sparticle species that could be detected at the LHC and
$e^+ e^-$ colliders with different centre-of-mass energies, in various benchmark supersymmetric
scenarios with non-universal Higgs masses ($\alpha, \beta, \gamma$) or gravitino dark matter
($\delta, \epsilon, \zeta, \eta$)~\protect\cite{Bench3}.}
\label{fig:Bench3}
\end{figure}

Each supersymmetric cascade at the LHC would terminate with a meta- stable stau, and
simulations show that these events could be selected with high efficiency. The stau mass
could then be determined by time-of-flight measurements with an accuracy below 
1~\%~\cite{Polesello,Are}.
Once one had identified the two staus in each supersymmetric event, one could
reconstruct the masses of the heavier sparticles that decay into them, e.g., $\chi \to
{\tilde \tau} \tau$ and $q_R \to q \chi$, as seen in Fig.~\ref{fig:Are}~\cite{Are}. 
As in the more familiar $\chi$ LSP case, large 
numbers of sparticles could be discovered in some supersymmetric scenarios, as seen
in Fig.~\ref{fig:Bench3}~\cite{Bench3}.

Some of the staus produced at the end-points of these cascades would be moving quite
slowly, and it may be possible to trap some of them as they slow down thanks to
conventional electromagnetic energy-loss mechanisms, either within an LHC detector
or in the surrounding material~\cite{Feng,Nojiri}. Very few of the staus would be travelling slow 
enough to
stop within the detector itself, but some of them might stop within a few metres of material.
There is no room to put much extra material inside either the ATLAS or CMS caverns,
except possibly in a forward direction where not many staus would be headed. However,
some staus would stop within the first few metres of concrete and rock surrounding the
cavern. The ATLAS or CMS muon system could probably be used to locate the impact
point on the cavern wall with a precision of a centimetre or so, and to fix the impact angle
with an accuracy $\sim 10^{-3}$. One might then be able to bore a hole into the cavern
wall and remove a core containing the stopped stau. Because of the radiation levels
during LHC operation, this could be done only during the two-day technical stops
scheduled each month, or during the long shutdowns each year~\cite{Bench3}. Therefore this
exercise would be useful only if the lifetimes exceeds about $10^6$~s.

This is just one example of the extra information about one particular exotic
new-physics scenario
that might be accessible to the LHC at its design luminosity.

\section{Extra Time}

\begin{figure}[htb]
\centerline{\includegraphics[height=2.5in]{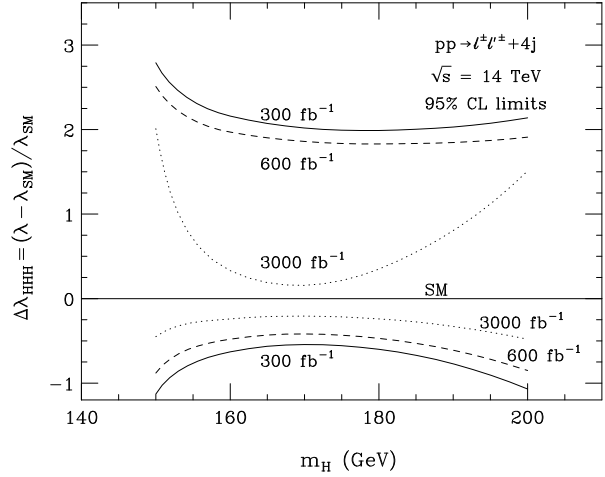}}
\caption{\it The expected sensitivities of the LHC and SLHC to the triple-Higgs 
coupling~\protect\cite{Baur}.}
\label{fig:3Higgs}
\end{figure}

\begin{figure}[htb]
\centerline{\includegraphics[height=3in]{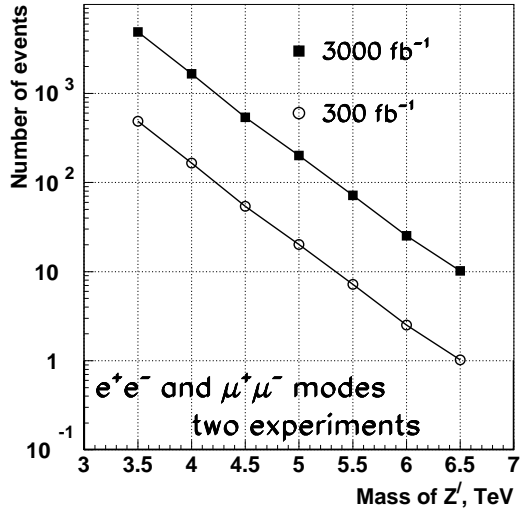}}
\caption{\it A comparison of the reaches of the standard LHC and the SLHC for a $Z'$ boson
with couplings similar to those of the $Z$ in the Standard Model~\protect\cite{SLHCphys}.}
\label{fig:extend}
\end{figure}

It appears technically possible to increase the LHC luminosity significantly beyond its
nominal design value of $10^{34}$~cm$^{-2}$s$^{-1}$, perhaps as high as
$10^{35}$~cm$^{-2}$s$^{-1}$, a possibility known as the SLHC~\cite{SLHCphys,PAF,POFPA}. 
This would make possible more sensitive studies of
a light Higgs boson and better searches for a very heavy Higgs boson, and might
provide the first sensitivity to the triple-Higgs coupling, as seen in 
Fig.~\ref{fig:3Higgs}~\cite{Baur}.
The SLHC would also make possible improved electroweak measurements, for
example of the triple-gauge-boson couplings, and extend the reaches for the searches for
heavy new physics. For example, it would extend significantly the reaches for
supersymmetry and new gauge bosons, as seen in Fig.~\ref{fig:extend}~\cite{SLHCphys}. If one were lucky, and supersymmetry or some other new physics such as extra dimensions had already appeared
during the nominal LHC running, the SLHC would also provide much greater statistics aand the
opportunity to explore this new physics in more detail.

However, increasing the LHC luminosity much beyond $10^{34}$~cm$^{-2}$s$^{-1}$
would require modifications to the
LHC interaction regions and to its injector chain.
It would also entail major modifications to the ATLAS and CMS detectors, principally the
inner tracking systems but also the calorimetry, trigger, and data-acquisition systems~\cite{POFPA}.
Overall, these might require efforts comparable to 30 to 50~\% of the original
investments in the major LHC detectors ATLAS and CMS.

\section{Penalty Shoot-out}

Who will go into the next round of our competition to understand
Nature? Will it be an $e^+ e^-$ collider, such as the ILC~\cite{Foster} or CLIC~\cite{CLIC}? 
Or will it be a higher-energy
hadron collider, such as an upgrade of the LHC energy to twice (DLHC) or three times
(TLHC) its current design value~\cite{POFPA}?

If the LHC discovers a lot of new low-mass physics, such as a light Higgs boson and
low-mass supersymmetry, there may be plenty of motivation for even a relatively low-energy
$e^+ e^-$ collider. However,  we cannot yet assume that a light Higgs boson necessarily
exists, and we do not even know whether suspersymmetry exists, let alone whether it
is light enough to be measured directly at the ILC. As seen in Fig.~\ref{fig:heavy}, the
range of sparticle masses allowed by our present ignorance extends well beyond the
reach of the ILC, even if one requires it to provide the dark matter. If supersymmetry does
lie within the ILC energy range, there will be interesting measurements to make, but even in
this case a higher-energy $e^+ e^-$ collider such as CLIC would be able to make and
measure the higher-mass sparticles in the supersymmetric spectrum. If sparticles exist beyond
the ILC range, CLIC may be even more essential. However, if sparticles are very heavy,
the DLHC or the TLHC may be more desirable projects.

Only the LHC will be able to tell us which, if any, of these possibilities is realized by Nature. But 
at least we may have the first answers quite soon. Already by 2010 the LHC should be providing 
key information about the Higgs boson and possible extensions of the Standard Model, on the
basis of which the planning for subsequent accelerators may become clearer.


\begin{thebibliography}{99}

\bibitem{Gonzalez} O. Gonzalez Lopes, Plenary talk at this meeting.

\bibitem{Hays} C. Hays, Plenary talk at this meeting.

\bibitem{Zielinski} M. Zielinski, Plenary talk at this meeting.

\bibitem{Bartoldus} R. Bartoldus, Plenary talk at this meeting.

\bibitem{Qian} J.-M. Qian, Plenary talk at this meeting.

\bibitem{Bigi} I. Bigi, plenary talk at this meeting.

\bibitem{Lykken} J. Lykken, Plenary talk at this meeting.

\bibitem{Harlander} R. Harlander, Plenary talk at this meeting.

\bibitem{Pokorski} S. Pokorski, Plenary talk at this meeting.

\bibitem{Kalinowski} J. Kalinowski, Plenary talk at this meeting.

\bibitem{hierarchy} L.~Maiani, {\it All You Need To Know About The Higgs Boson}, Proceedings
of the Gif-sur-Yvette Summer School On Particle Physics, 1979, pp.1-52; G.~'t~Hooft, in {\it
Recent developments in Gauge Theories}, Proceedings of the NATO Advanced Study
Institute, Carg{\`e}se, 1979, eds. G.~'t~Hooft et al. (Plenum Press, NY, 1980);
E.~Witten,
  Phys.\ Lett.\ B {\bf 105} (1981) 267.

\bibitem{unification} J.~R.~Ellis, S.~Kelley and D.~V.~Nanopoulos,
  Phys.\ Lett.\ B {\bf 249} (1990) 441 and
  Phys.\ Lett.\ B {\bf 260} (1991) 131;
U.~Amaldi, W.~de Boer and H.~Furstenau,
  Phys.\ Lett.\ B {\bf 260} (1991) 447;
  C.~Giunti, C.~W.~Kim and U.~W.~Lee,
  Mod.\ Phys.\ Lett.\ A {\bf 6} (1991) 1745.
  
\bibitem{higgs} J.~R.~Ellis, D.~Nanopoulos, K.~A.~Olive and Y.~Santoso,
  Phys.\ Lett.\ B {\bf 633} (2006) 583
  [arXiv:hep-ph/0509331].

\bibitem{Erler} J.~Erler, {\it Fit to electroweak precision data},
  arXiv:hep-ph/0606148.

\bibitem{EHNOS} J.~R.~Ellis, J.~S.~Hagelin, D.~V.~Nanopoulos, K.~A.~Olive and M.~Srednicki,
  Nucl.\ Phys.\ B {\bf 238} (1984) 453.

\bibitem{EOSS} J.~R.~Ellis, K.~A.~Olive, Y.~Santoso and V.~C.~Spanos,
  Phys.\ Lett.\ B {\bf 603} (2004) 51
  [arXiv:hep-ph/0408118].

\bibitem{EOS} J.~R.~Ellis, K.~A.~Olive and Y.~Santoso,
  New J.\ Phys.\  {\bf 4} (2002) 32
  [arXiv:hep-ph/0202110].

\bibitem{EHOW} J.~Ellis, S.~Heinemeyer, K.~A.~Olive and G.~Weiglein,
{\it Indications of the CMSSM mass scale from precision electroweak data},
  arXiv:hep-ph/0604180.

\bibitem{Engelen} J. Engelen, Plenary talk at this meeting.

\bibitem{Jenni} P. Jenni, Plenary talk at this meeting.

\bibitem{Dobrzynski} L. Dobrzynski, Plenary talk at this meeting.

\bibitem{Boonekamp} M. Boonekamp, Plenary talk at this meeting.

\bibitem{Unal} G. Unal, Plenary talk at this meeting.

\bibitem{POFPA}
A.~Blondel, L.~Camilleri, A.~Ceccucci, J.~Ellis, M.~Lindroos, M.~Mangano and G.~Rolandi,
{\it Physics opportunities with future proton accelerators at CERN},
  arXiv:hep-ph/0609102.
  
\bibitem{Spiropulu} M. Spiropulu, Plenary talk at this meeting.

\bibitem{Salgado} C. Salgado, Plenary talk at this meeting.

\bibitem{Roland} G. Roland, Plenary talk at this meeting.

\bibitem{CGC} E.~Iancu, A.~Leonidov and L.~D.~McLerran,
  Nucl.\ Phys.\ A {\bf 692} (2001) 583
  [arXiv:hep-ph/0011241].

\bibitem{AdSCFT} 
P.~Kovtun, D.~T.~Son and A.~O.~Starinets,
  Phys.\ Rev.\ Lett.\  {\bf 94} (2005) 111601
  [arXiv:hep-th/0405231].

\bibitem{Nappi} E. Nappi, Plenary talk at this meeting.

\bibitem{Nakada} T. Nakada, Plenary talk at this meeting.

\bibitem{Bartels} J. Bartels, Plenary talk at this meeting.

\bibitem{deRoeck} A. de Roeck, Plenary talk at this meeting.

\bibitem{Avati} V. Avati, Plenary talk at this meeting.

\bibitem{Higgsspin}
S.~Y.~Choi, D.~J.~Miller, M.~M.~Muhlleitner and P.~M.~Zerwas,
  Phys.\ Lett.\ B {\bf 553} (2003) 61
  [arXiv:hep-ph/0210077];
C.~P.~Buszello, I.~Fleck, P.~Marquard and J.~J.~van der Bij,
  Eur.\ Phys.\ J.\ C {\bf 32} (2004) 209
  [arXiv:hep-ph/0212396].

\bibitem{HiggsCP} E.~Accomando {\it et al.},
{\it Workshop on CP studies and non-standard Higgs physics},
  arXiv:hep-ph/0608079.

\bibitem{Bench1} M.~Battaglia {\it et al.},
  Eur.\ Phys.\ J.\ C {\bf 22} (2001) 535
  [arXiv:hep-ph/0106204].

\bibitem{Bench2} M.~Battaglia, A.~De Roeck, J.~R.~Ellis, F.~Gianotti, K.~A.~Olive and L.~Pape,
  Eur.\ Phys.\ J.\ C {\bf 33} (2004) 273
  [arXiv:hep-ph/0306219].

\bibitem{Bench3} A.~De Roeck, J.~R.~Ellis, F.~Gianotti, F.~Moortgat, K.~A.~Olive and L.~Pape,
  arXiv:hep-ph/0508198.

\bibitem{Are} J.~R.~Ellis, A.~R.~Raklev and O.~K.~Oye,
  JHEP {\bf 0610} (2006) 061
  [arXiv:hep-ph/0607261].

\bibitem{GDM} J.~R.~Ellis, K.~A.~Olive, Y.~Santoso and V.~C.~Spanos,
  Phys.\ Lett.\ B {\bf 588} (2004) 7
  [arXiv:hep-ph/0312262], and
Phys.\ Rev.\ D {\bf 70} (2004) 055005
  [arXiv:hep-ph/0405110].
  
\bibitem{EOV} J.~R.~Ellis, K.~A.~Olive and E.~Vangioni,
  Phys.\ Lett.\ B {\bf 619} (2005) 30
  [arXiv:astro-ph/0503023].

\bibitem{CEFOS} R.~H.~Cyburt, J.~Ellis, B.~D.~Fields, K.~A.~Olive and V.~C.~Spanos,
  arXiv:astro-ph/0608562.

\bibitem{Polesello} S.~Ambrosanio, B.~Mele, A.~Nisati, S.~Petrarca, G.~Polesello, A.~Rimoldi and G.~Salvini,
  arXiv:hep-ph/0012192.

\bibitem{Feng} J.~L.~Feng and B.~T.~Smith,
  Phys.\ Rev.\ D {\bf 71} (2005) 015004
  [Erratum-ibid.\ D {\bf 71} (2005) 0109904]
  [arXiv:hep-ph/0409278].

\bibitem{Nojiri} K.~Hamaguchi, Y.~Kuno, T.~Nakaya and M.~M.~Nojiri,
  Phys.\ Rev.\ D {\bf 70} (2004) 115007
  [arXiv:hep-ph/0409248].

\bibitem{SLHCphys} F.~Gianotti {\it et al.},
  Eur.\ Phys.\ J.\ C {\bf 39} (2005) 293
  [arXiv:hep-ph/0204087].

\bibitem{PAF}
M.~Benedikt, R.~Garoby, F.~Ruggiero, R.~Ostojic, W.~Scandale, E.~Shaposhnikova and J.~Wenninger, {\it Preliminary accelerator plans for maximizing the integrated LHC luminosity},
CERN-AB-2006-018.

\bibitem{Baur}
U.~Baur, T.~Plehn and D.~L.~Rainwater,
  Phys.\ Rev.\ D {\bf 67} (2003) 033003
  [arXiv:hep-ph/0211224].

\bibitem{Foster} B. Foster, Plenary talk at this meeting.

\bibitem{CLIC}
E.~Accomando {\it et al.}  [CLIC Physics Working Group],
{\it Physics at the CLIC multi-TeV linear collider},
  arXiv:hep-ph/0412251.

\end{thebibliography}
\end{document}